\documentclass[letter,twocolumn]{jpsj3}
\usepackage{txfonts}
\usepackage[dvipdfmx]{graphicx} 
\usepackage{dcolumn}
\usepackage{bm}
\usepackage{amsmath}
\usepackage{braket}
\usepackage{amsfonts}
\usepackage{amssymb}
\usepackage{color}

\newcommand{\beq}{\begin{equation}}
\newcommand{\eeq}{\end{equation}}

\title{Noise-Tolerance of Majorana Teleportation \\in Mesoscopic Topological Superconductors}

\author{Tsukasa Goto\thanks{goto@blade.mp.es.osaka-u.ac.jp}, Masayuki Sugeta, Takeshi Mizushima, and Satoshi Fujimoto}
\inst{Department of Materials Engineering Science, Osaka University, Toyonaka 560-8531, Japan} 

\abst{
We investigate teleportation interference associated with the non-local character of Majorana zero modes (MZMs) as a probe of MZMs focusing on the tolerance of teleportation against disturbances, such as inhomogeneous potentials at junctions and disorder. 
We develop a method for calculating non-local conductance in mesoscopic topological superconductors with fixed parity. In the trivial phase, the non-local conductance exhibits the $h/2e$-periodicity, while in the topological phase with fixed parity, it exhibits the $h/e$-periodicity, indicative of Majorana teleportation. We find that the $h/e$-periodicity is stable against changes in inhomogeneous potential structures and disorder. 
These results imply that MZMs can cause teleportation interference even in the presence of disturbances, leading to a clear distinction between the trivial and topological phases.
}

\begin{document}
\maketitle
Topological superconductors host zero-energy mid-gap states called the Majorana zero modes (MZMs), 
which have been attracting much attention for their potential applications in fault-tolerant topological quantum computation\cite{nayakNonAbelianAnyonsTopological2008, sarmaMajoranaZeroModes2015}.
However, despite various experimental methods, the experimental detection of MZMs is yet to be established\cite{yazdaniHuntingMajoranas2023, zhangNextStepsQuantum2019,lut18,pra20,agh23}.
For example, local conductance measurements have been used to experimentally detect MZMs~\cite{mou12,das12,den12,fin13,den16,nicheleScalingMajoranaZeroBias2017,che17,gul18,zha21,aliceaNewDirectionsPursuit2012,lut18,pra20,mar22}, 
as perfect Andreev reflections mediated via MZMs cause quantized zero bias conductance peaks (ZBCPs)~\cite{akh11,ful11,aliceaNewDirectionsPursuit2012,mar22}.
However, it has been reported that even a trivial phase can realize nearly quantized ZBCPs\cite{kel12,pra12,roy13,sta14,cay15,liu17,mooreTwoterminalChargeTunneling2018, mooreQuantizedZerobiasConductance2018} because the presence of inhomogeneous potentials near the junctions between a superconductor and an electrode generate low-energy trivial bound states. These bound states, referred to as partially-separated Andreev bound states (ps-ABSs), mimic the local properties of MZMs. Therefore, the observation of quantized ZBCPs does not represent sufficient evidence for the existence of MZMs, motivating the exploration of other experimental schemes to detect MZMs. 

Teleportation interferometry associated with two MZMs is a possible alternative tool for this purpose\cite{fuElectronTeleportationMajorana2010, zazunovCoulombBlockadeMajoranafermioninduced2011, whiticarCoherentTransportMajorana2020}. 
It utilizes non-local characters of MZMs arising from fractionalization of electrons, which cannot be probed by local measurements.
This teleportation leads to the transmission of a phase-coherent single electron through MZMs, independent of the distance between the MZMs. It is noted that the teleportation interference is also an effective tool for readout of topological qubits\cite{pluggeRoadmapMajoranaSurface2016, vijayTeleportationbasedQuantumInformation2016}.
However, the influence of trivial bound states, such as ps-ABSs, on this scheme has not been well understood.
This is a nontrivial issue because ps-ABSs at junctions can hybridize with MZMs, which may destroy the coherence of two spatially separated MZMs.
In fact, such hybridization processes disturb non-Abelian braiding dynamics of MZMs \cite{tanaka}.
Thus, it is highly desirable to clarify the tolerance of Majorana teleportation against the existence of ps-ABSs.
This issue is also technically challenging due to difficulties in conserving fermion parity while calculating transport coefficients, which is necessary for Majorana teleportation interference.
The conservation of electron number parity can be ensured by applying electrostatic energy to mesoscopic superconductors\cite{fuElectronTeleportationMajorana2010}.
For the implementation of this effect in {\it ab initio} numerical calculations, it is necessary to address quantum many-body problems involving Coulomb interactions between electrons, which requires high computational costs.

In this paper, to overcome this issue, we develop an approach of a parity-fixed Green's function method and investigate the effects of noises such as ps-ABSs and disorder on Majorana teleportation with the use of {\it ab initio} numerical calculations, which do not presume the existence of MZMs.
We find that non-local conductance characterizing the teleportation interference shows clearly distinct behaviors between the trivial and topological phases, even under inhomogeneous potentials that induce ps-ABSs, providing strong evidence of the teleportation effect due to MZMs. 
This effect remains stable regardless of the potential structure and disorder effects, implying that teleportation interference can be a viable experimental scheme for detecting MZMs.

\begin{figure}[t]
  \includegraphics[width=\columnwidth]{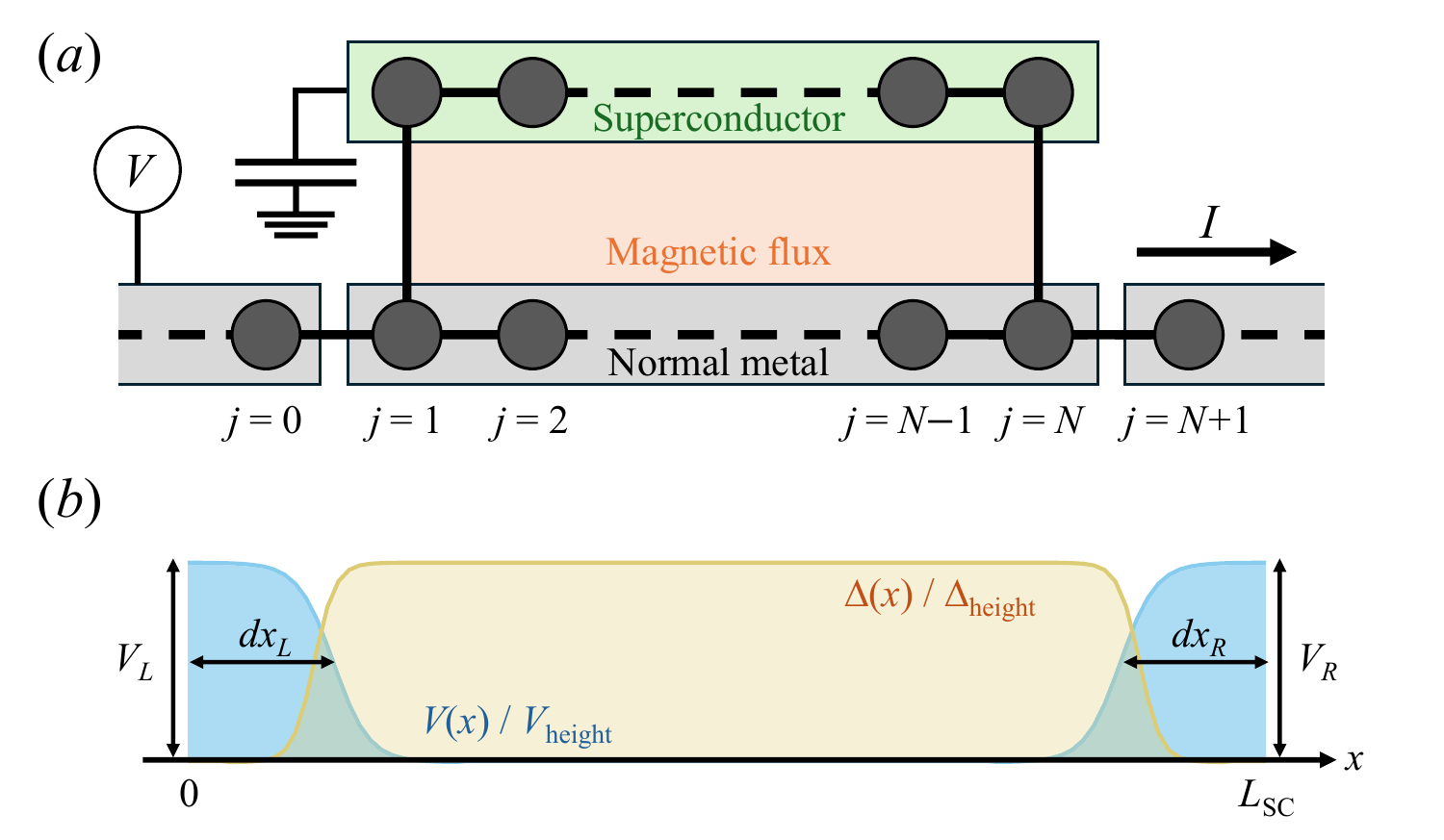}
  \caption{
    (\textit a) Schematic image of our setup. The metallic and superconducting nanowires are connected at both ends, forming an SN loop system. 
    In addition, semi-infinite metal leads are connected to the metal part of the loop.
    (\textit b) Spatial distribution of inhomogeneous potentials of the topological superconductor with the junctions at the ends.
  }
  \label{fig1}
\end{figure}
The setup considered in this paper is shown in Fig.\ref{fig1}(\textit a).
A semiconductor nanowire with the Rashba spin-orbit interaction can be effectively treated as a topological superconductor by adding the Zeeman effect and the proximity effect due to $s$-wave superconductors\cite{satoNonAbelianTopologicalOrder2009,lutchynMajoranaFermionsTopological2010,oregHelicalLiquidsMajorana2010}.
By connecting the ends of the superconducting (SC) and normal metal (N) nanowires together, the system becomes a one-dimensional SN-loop system, which is described by the Hamiltonian
$\mathcal{H}_{\rm{loop}} = \mathcal{H}_{\rm{SC}}+\mathcal{H}_{\rm{N}}+\mathcal{H}_{\rm{hop-SN}}$ as,
\begin{align}
  \mathcal{H}_{\rm{SC}} = &\sum_{j,\sigma}(-t c_{j+1,\sigma}^{\dagger}c_{j,\sigma}+\rm{H.c.}) \notag \\
          &+ \sum_{j,\sigma} \left(V(x_{j})-\mu_{\rm{SC}}\right)c_{j,\sigma}^{\dagger}c_{j,\sigma} + h(c_{j,\uparrow}^{\dagger}c_{j,\uparrow}- c_{j,\downarrow}^{\dagger}c_{j,\downarrow})  \notag \\
          &+ \sum_{j}(-\lambda c_{j-1,\downarrow}^{\dagger}c_{j,\uparrow} + \lambda c_{j+1,\downarrow}^{\dagger}c_{j,\uparrow} + \rm{H.c.})  \notag \\
          &+ \sum_{j}(\Delta(x_{j})c_{j,\uparrow}^{\dagger} c_{j,\downarrow}^{\dagger} + \rm{H.c.}),
\end{align}
\begin{align}
  \label{metal}
  \mathcal{H}_{\rm{N}} = \sum_{j,\sigma}(-t\psi_{j+1,\sigma}^{\dagger} \psi_{j,\sigma}+\rm{H.c.}) -\mu_{\rm{N}} \psi_{j,\sigma}^{\dagger}\psi_{j,\sigma},
\end{align}
\begin{align}
  \mathcal{H}_{\rm{hop-SN}} = \sum_{\sigma} \left[-t_{\rm{SN}}(c_{1,\sigma}^{\dagger} \psi_{1,\sigma} + e^{i\phi/2}c_{N,\sigma}^{\dagger} \psi_{N,\sigma}) +\rm{H.c.} \right],
  \label{hopSN}
\end{align}
where $c_{j,\sigma}^{\dagger} (c_{j,\sigma})$, $\psi_{j,\sigma}^{\dagger}(\psi_{j,\sigma})$ denotes a creation (annihilation) operator of electrons with spin $\sigma =(\uparrow, \downarrow)$ at site $j$ in the 
superconductor region and the metal region, respectively, 
and $t_{\rm{S},\rm{N},\rm{SN}}$, $\mu_{\rm{S},\rm{N}}$, $h$, $\lambda$ are a hopping amplitude between nearest-neighboring sites where the subscripts S, N and SN denote the region of superconductor, normal metal and SN junctions,
a chemical potential, the Zeeman field, and the Rashba interaction strength, respectively.
In Eq.~\eqref{hopSN}, which represents hopping at SN junctions, a Peierls phase factor $e^{i\phi /2}$ is introduced, which depends on the magnetic flux penetrating the one-dimensional loop system.
Furthermore, we consider the inhomogeneous potentials stemming from the quantum dot and the pair potentail\cite{mooreQuantizedZerobiasConductance2018, sugetaEnhanced2pperiodicAharonov2023} that create ps-ABSs as shown in Fig.~\ref{fig1}(\textit b). The quantum dot potential is given by,
\begin{align}
  V(x) = &\frac{V_{\rm L}}{2} \left[1- \tanh \left(\frac{x-dx_{\rm L}}{\delta x_{V}}\right)\right] \notag \\
  &+ \frac{V_{\rm R}}{2} \left[1- \tanh \left(\frac{x-L_{\rm{SC}}+dx_{\rm R}}{\delta x_{V}}\right)\right], 
\end{align}
where $V_{\rm L,R}$ and $dx_{\rm L,R}$ are the height and width of the quantum dot potential respectively, $\delta x_{V}$ is the length scale over which $V(x)$ varies, and the subscripts L and R denote the respective parameters on the $x=0, x=L_{\rm{SN}}$ side.
The pair potential is given by,
\begin{align}
  \Delta(x) = &\frac{\Delta_{0}}{2} \left[1- \tanh \left(\frac{x-dx_{\rm{SC}}}{\delta x_{\Delta}}\right)\right] \notag \\
  &\times \frac{1}{2} \left[1- \tanh \left(\frac{x-L_{\rm{SC}}+dx_{\rm{SC}}}{\delta x_{\Delta}}\right)\right],
\end{align}
where $\Delta_{0}$ is the height of the pair potential, $dx_{\rm{SC}}$ is a parameter that controls the extension of the pair potential in the quantum dot region due to the proximity effect and $\delta x_{\Delta}$ is the length scale over which $\Delta (x)$ varies.
For the calculations of non-local conductance, we consider a system of semi-infinite metal leads connected at both ends of the metal part of the SN-loop system. Here, the Hamiltonian of the semi-infinite metal leads is also the same as in Eq.~\eqref{metal}.
Throughout this paper, we set parameters $t_{\rm{SC}}=t_{\rm{N}}=t$, $t_{\rm{SN}}=0.1t$, $\mu_{\rm{SC}}=-1.975t$, $\mu_{\rm{N}}=0$, $\lambda=0.3t$, $\Delta_{0}=0.1t$, $V_{L}=V_{R}=0.35\Delta_{0}$, $dx_{L}=dx_{R}=20.5a$, $\delta_{V}=5a$, $dx_{\rm{SC}}=17.8a$, $\delta_{\Delta}=3a$. The length of the nanowire is $L_{\rm{SC}}=L_{\rm{N}}=500a$ unless stated otherwise, where $a$ is the lattice constant of the tight-binding model.

The Lee-Fisher formula\cite{fisherRelationConductivityTransmission1981, leeAndersonLocalizationTwo1981, wakatsukiFermionFractionalizationMajorana2014} is used for the calculation of the non-local conductance,
\begin{align}
  \label{Lee-Fisher}
  \Gamma_{ij} = \frac{2e^2}{h}t^2\rm{Tr}\bigg[ P_{e} \Big(&\tilde{G}_{i,j}\tilde{G}_{j+1,i+1} + \tilde{G}_{i+1,j+1}\tilde{G}_{j,i} \notag \\
  & -\tilde{G}_{i,j+1}\tilde{G}_{j,i+1} - \tilde{G}_{i+1,j}\tilde{G}_{j+1,i}\Big)\bigg],
\end{align}
where $\tilde{G}_{i,j} = (G^{\rm R}_{i,j}-G^{\rm A}_{i,j})/2i$ with the retarded (advanced) Green's function $G^{\rm R}_{i,j}$ ($G^{\rm A}_{i,j}$) and $P_{e}$ is the projection operator onto a sector of the electron number parity.
We obtain the parity-fixed Green's function in the following three steps. 
First, we diagonalize the Hamiltonian of the SN loop system ($\mathcal{H}_{\rm loop}$) and separate the energy spectrums into two parity sectors using the flux dependence of the eigenvalues. The eigenstates exhibit an $h/e$-periodicity with respect to the magnetic flux in the case with MZMs, and a phase difference of $h/2e$ between even and odd parity sectors\cite{kitaevUnpairedMajoranaFermions2001, kwonFractionalAcJosephson2004}.
From the eigenstates in each parity sectors, the Green's function with fixed parity of the loop system is constructed as,
\begin{align}
  \label{Green}
  G_{{\rm R}/{\rm A}}^{\rm{parity}} =  \frac{\ket{\phi_{\rm{parity}}}\bra{\phi_{\rm{parity}}}}{E-\epsilon_{\rm{parity}} \pm i\delta} + \sideset{}{'}{\sum}_{|\epsilon|<\epsilon_{c}} \frac{\Ket{\phi_{\epsilon}}\Bra{\phi_{\epsilon}}}{E-\epsilon \pm i\delta},
\end{align}
where the first term in Eq.~\eqref{Green} represents the contribution of the MZMs, and $\ket{\phi_{\rm{parity}}}$ is the eigenstate for a MZM in the even or odd parity sector. 
The second term in Eq.~\eqref{Green} represents the contribution of the other modes. We introduce a cut-off energy for the summation in the second term because the Lee-Fisher formula is formulated at sufficiently low temperatures.\cite{leeAndersonLocalizationTwo1981}.
Second, the Green's functions of the left and right semi-infinite metal leads are obtained by using the recursive Green's function method\cite{umerskiClosedformSolutionsSurface1997,lewenkopfRecursiveGreenFunction2013}.
Finally, the junctions between the metal part of the SN loop system and the semi-infinite leads are taken into account, and the total Green's function $G$ is obtained from the Dyson formula as,\cite{lewenkopfRecursiveGreenFunction2013}
\begin{align}
  &G = G^{(0)} + G^{(0)}VG , \\
  &G^{(0)} = g_{0}+g^{\rm{SN}}+g_{N+1} , \\
  &V = U_{0,1}+U_{1,0}+U_{N,N+1}+U_{N+1,N}, 
\end{align}
where $g^{\rm{SN}}$ is the Green's functions of the SN-loop system, and $g_{0}$, $g_{N+1}$ are, respectively, the surface Green's functions of the left and right semi-infinite leads.
In the calculation of non-local conductance, Eq.~\eqref{Lee-Fisher}, we set $i=0$, $j=N$.
Therefore, we need to find the $(0,N)$, $(0,N+1)$, $(1,N)$, $(1,N+1)$, $(N,0)$, $(N,1)$,$(N+1,0)$ and $(N+1,1)$ components of 
the Green's function matrix $G$.
Here, the matrix components are given by,
\begin{align}
  \label{component}
  G_{i,j} = \Bra{i}G\Ket{j} = \Bra{i}G^{(0)}\Ket{j} + \Bra{i}G^{(0)}VG \Ket{j}. 
\end{align}
The total Green's function is obtained by calculating Eq.~\eqref{component} for each component. We can then obtain a total Green's function with fixed parity and compute a fixed parity non-local conductance from Eq.~\eqref{Lee-Fisher}.

\begin{figure}[t!]
  \includegraphics[width=\columnwidth]{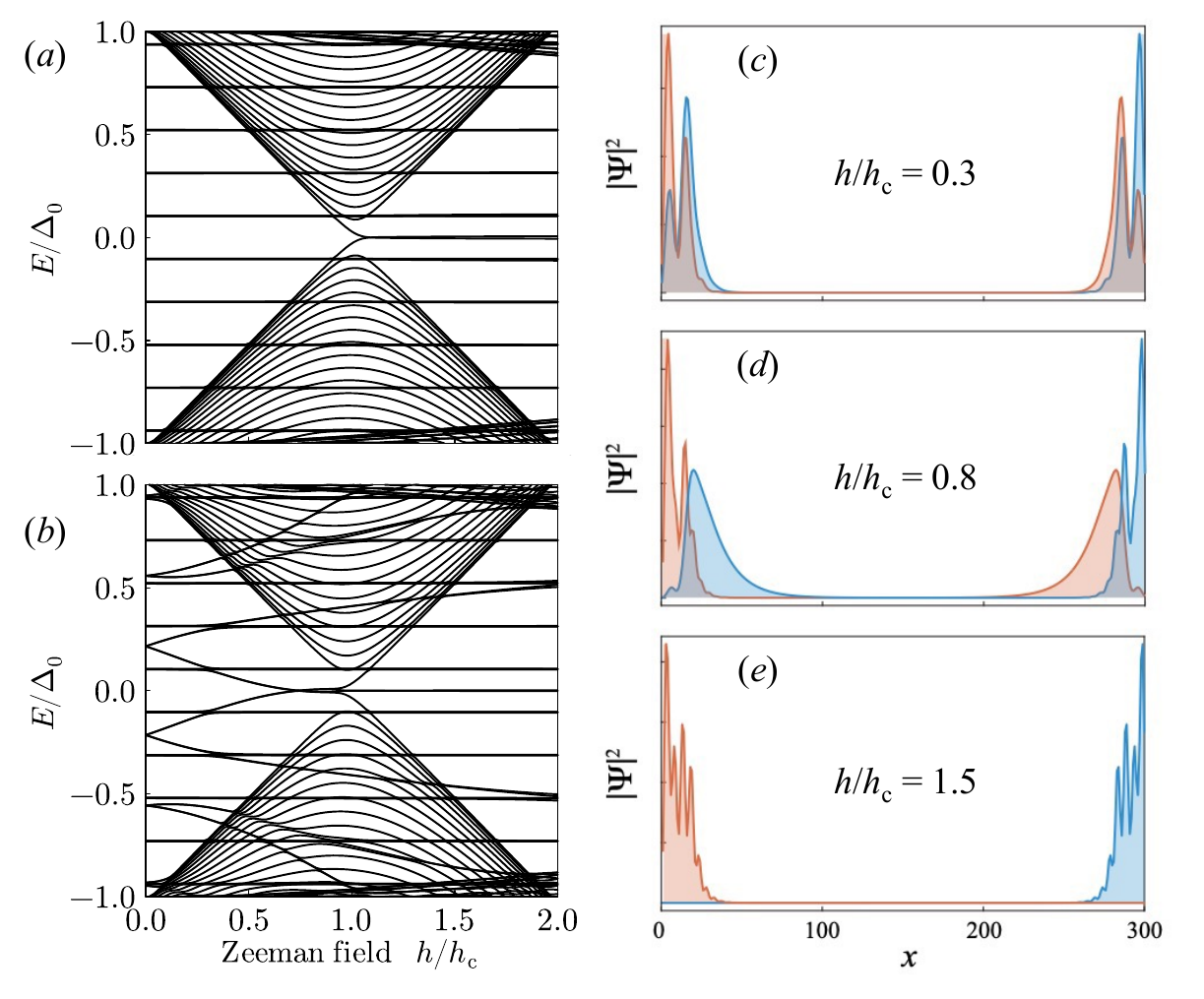}
  \caption{
    (\textit a-\textit b) Magnetic field dependence of eigenvalues of the SN loop system. In (\textit a) there are no inhomogeneous potentials ($V(x)=0, \Delta(x)=\Delta_{0}$), while (\textit b) shows the case where there are.
    (\textit c-\textit e) represent the low-energy spatial distribution of the wave function for (\textit c) ABSs ($h/h_{c}=0.3$), (\textit d) ps-ABSs ($h/h_{c}=0.8$) and  (\textit e) MZMs ($h/h_{c}=1.5$) respectively. $L_{\rm{SC}}=L_{\rm{N}}=300a$.
  }
  \label{fig2}
\end{figure}
To investigate the effects of ps-ABSs on Majorana teleportation, we first clarify the magnetic field regime in which ps-ABSs are realized.
Figures~\ref{fig2}(\textit a) and \ref{fig2}(\textit b) show the magnetic flux dependence of the eigenenergy in the SN loop system. Here, $h_{\rm c}$ is the critical magnetic field separating the trivial and topological phases.
The region of $h<h_{\rm c}$ corresponds to the trivial phase with a finite energy gap, while $h>h_{\rm c}$ is the topological phase with MZMs. In the case of no inhomogeneous potential [Fig.~\ref{fig2}(\textit a)], the energy gap closes at $h=h_{\rm c}$, showing a clear signature of the topological phase transition. However, the existence of inhomogeneous potential results in the formation of nearly zero-energy edge states even in the trivial phase [Fig.\ref{fig2}(\textit b)].
By diagonalizing $\mathcal{H}_{\rm loop}$, we obtain the spatial distribution of bound states around the ends of the superconducting region,
\begin{align}
  \label{distribution}
  \left| \Psi_{\pm}(x) \right|^{2}  = \left| \frac{\phi_{\epsilon}(x) \pm \phi_{-\epsilon}(x)}{\sqrt{2}} \right|^{2},
\end{align}
where $\phi_\epsilon$ is the eigenstate with the eigenenergy $\epsilon$. Figures~\ref{fig2}(\textit c), \ref{fig2}(\textit d), and \ref{fig2}(\textit e) show the spatial distribution of the wave functions of the ABSs, ps-ABSs, and MZMs, respectively\cite{mooreTwoterminalChargeTunneling2018}. The low-energy modes are localized at both ends in the case of MZMs [Fig.\ref{fig2}(\textit e)], while the two modes overlap for the conventional ABSs [Fig.\ref{fig2}(\textit c)]. The ps-ABSs are realized in $0.75 \le h/h_{c} \le 1$, where the two modes are partially separated as shown in Fig.~\ref{fig2}(\textit d).

\begin{figure}[t!]
  \includegraphics[width=\columnwidth]{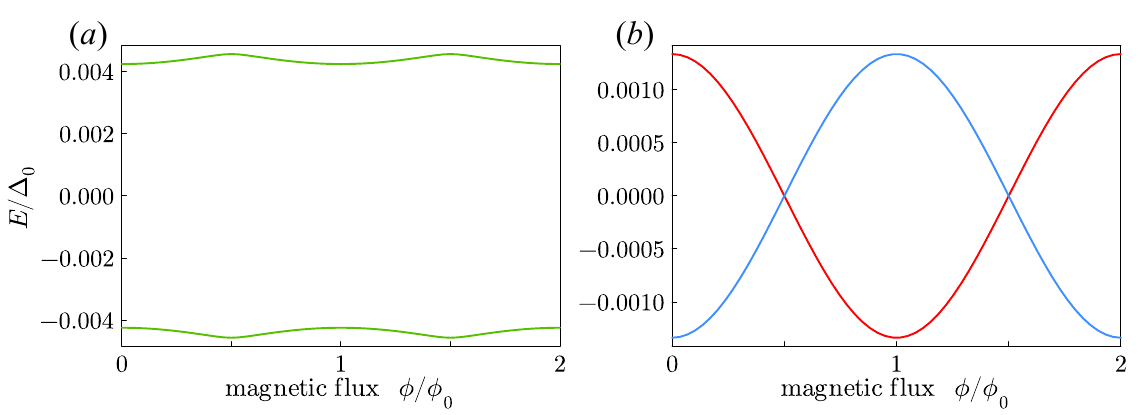}
  \caption{
    Magnetic flux dependence of the eigenenergy of the SN loop system for  (\textit a) ps-ABSs ($h/h_{c}=0.8$), and (\textit b) MZMs ($h/h_{c}=1.5$). 
    In (\textit b),  the red and blue lines represent cases of even and odd parity, respectively. $\phi_{0}$ is the superconducting magnetic flux quantum $h/2e$.
  }
  \label{fig3}
\end{figure}

\begin{figure}[t!]
  \includegraphics[width=\columnwidth]{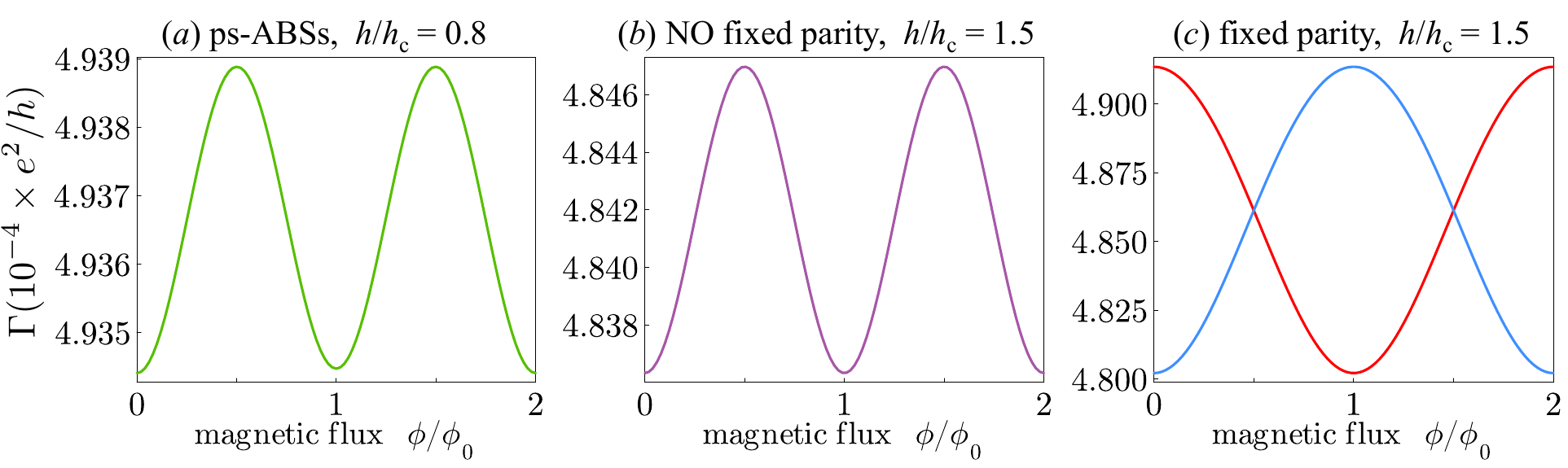}
  \caption{
    Magnetic flux dependence of the non-local conductance for (\textit a) the trivial phase ($h/h_{c}=0.8$), (\textit b) the topological phase ($h/h_{c}=1.5$) without fixed parity, and  (\textit c) the topological phase ($h/h_{c}=1.5$) with fixed parity, where $E=0$, $\delta=10^{-2}$, $\epsilon_{c}/\Delta_{0}=0.25$ in Eq.~\eqref{Green}. 
    In (\textit c), the red and blue lines represent cases of even and odd parity, respectively.
  }
  \label{fig4}
\end{figure}

We now examine the flux-dependence of the energy eigenvalues for these states.
As shown in Fig.~\ref{fig3}(\textit a), the ps-ABSs in the trivial phase exhibit the $h/2e$-periodicity. In contrast, Fig.~\ref{fig3}(\textit b) shows the eigenenergies in the topological phase, which exhibit the $h/e$-periodicity. The phase difference in Fig.~\ref{fig3}(\textit b) reflects the electron number parity associated with the MZMs.
We utilize these parity-resolved eigenstates to construct the Green's function in Eq.~\eqref{Green} and calculate the non-local conductance from Eq.~\eqref{Lee-Fisher}.
Figure~\ref{fig4} shows the magnetic field dependence of the non-local conductance.
It is found that, as shown in Fig.\ref{fig4}(\textit a), the non-local conductance exhibits  $h/2e$-periodicity in the trivial phase. Similarly, in the topological phase with non-fixed parity [Fig.\ref{fig4}(\textit b)], the non-local conductance also shows $h/2e$-periodicity
In the topological phase with fixed parity [Fig.\ref{fig4}(\textit c)], however, it exhibits a periodicity of $h/e$ with a phase difference of $h/2e$ between even and odd parity sectors. 
These results imply that even in the presence of inhomogeneous potentials, MZMs can cause teleportation interference in the topological phase while maintaining parity conservation. Furthermore, the distinct periodicity of the trivial and topological phases enables a clear distinction between MZMs and ps-ABSs.

\begin{figure*}[t!]
  \includegraphics[width=180mm]{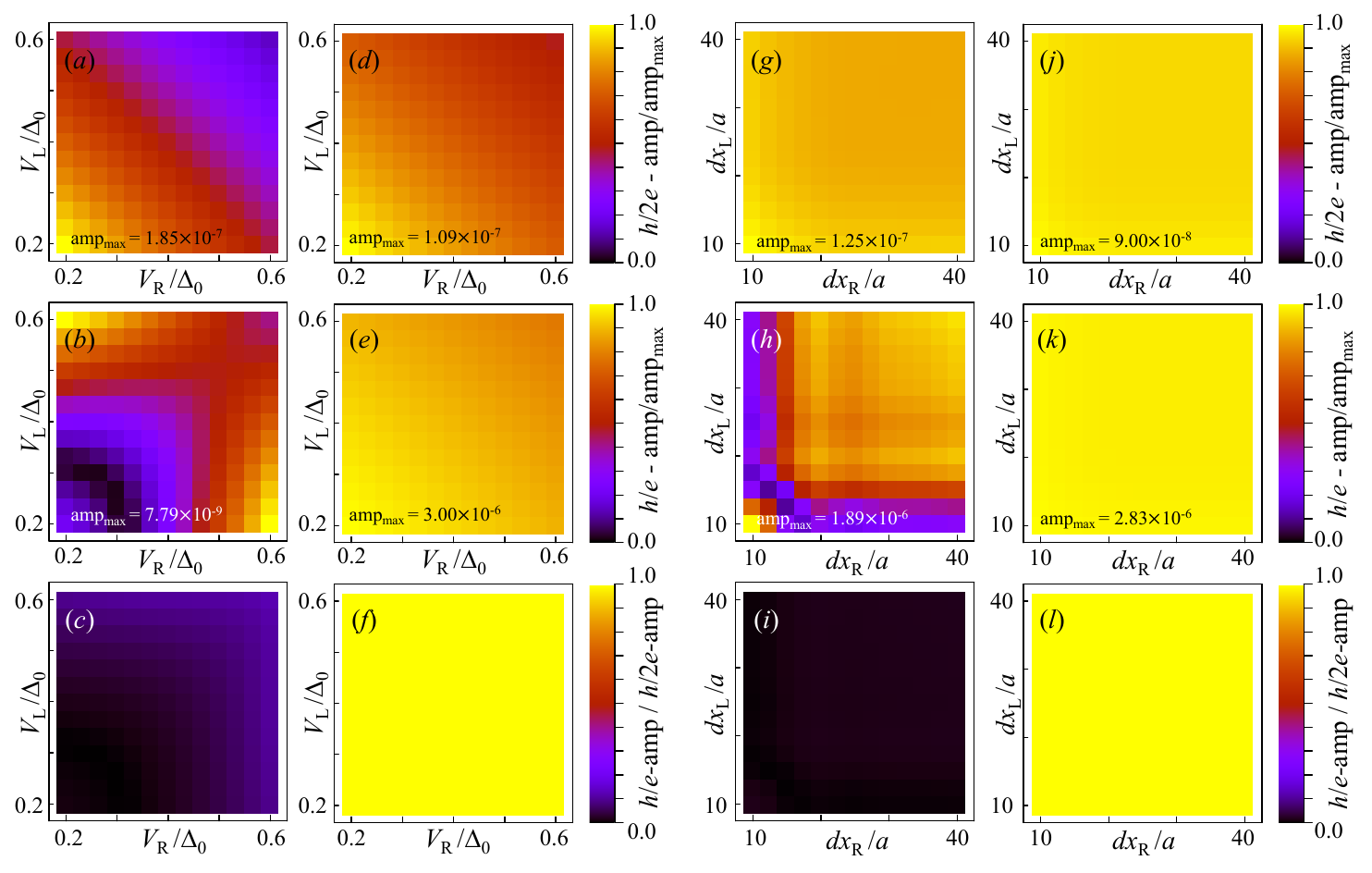}
  \caption{
    The amplitudes of the Fourier-transformed non-local conductance as functions of potential height (\textit a-\textit f) and width (\textit g-\textit l) at the left and right ends shown in Fig.~\ref{fig1}(\textit b).
    The top, middle, and bottom rows show the amplitudes with $h/2e$-periodicity, the amplitudes with $h/e$-periodicity, and 
    the ratio of the $h/e$-amplitude to the $h/2e$-amplitude, respectively.
    In  (\textit a-\textit c) and (\textit g-\textit i), the results for the trivail phase with $h/h_{c}=0.8$ are shown. In (\textit d-\textit f) and (\textit j-\textit l), the results for the topological phase with $h/h_{c}=1.5$ are shown.
  }
  \label{fig5}
\end{figure*}
We investigate the stability of Majorana teleportation by varying the height and width of the inhomogeneous potentials shown in Fig.~\ref{fig1}(\textit b). 
For this purpose, we perform the Fourier transformation of the non-local conductance,
\begin{align}
  \label{Fourier}
  \Gamma(\Phi) = \sum_{\omega} e^{i\omega \Phi} \Gamma_{\omega},
\end{align}
where $\Phi = 2\pi \phi/\phi_{0}$ and $\omega=n/2, n\in \mathbb{Z}$. The Fourier components, 
$\Gamma_{1/2}$ and $\Gamma_{1}$, correspond to the amplitudes with $h/e$ and $h/2e$-periodicity, respectively.

To examine the stability of teleportation interference, in Fig.~\ref{fig5}, we plot the amplitudes of $\Gamma_{1/2}$ and $\Gamma_{1}$ as functions of potentail parameters. In Fig.~\ref{fig5}(\textit a-\textit f), the $h/2e$-amplitude is larger in the trivial phase (Fig.~\ref{fig5}(\textit c)), whereas the $h/e$-amplitude dominates in the topological phase [Fig.\ref{fig5}(\textit f)]. 
Figures~\ref{fig5}(\textit a,\textit b) and \ref{fig5}(\textit d,\textit e) show that the $h/e$-periodicity in the topological phase is substantially more stable than in the trivial phase when the potential height is modified. Moreover, the maximum value of the $h/e$ amplitude in the topological phase is three orders of magnitude larger than the maximum value of the $h/2e$ amplitude in the trivial phase.
Majorana teleportation is stable also against the change of the potential width, as seen in Fig.~\ref{fig5}(\textit g-\textit l).
These results demonstrate that Majorana teleportation is significantly robust against inhomogeneous potentials at junctions, making it a distinct signature of MZMs in realistic junction systems.
We have also examined effects of impurity potentials, and confirmed the above results are not qualitatively affected by disorder. \cite{suppl}

In summary, we have demonstrated that the teleportation interference mediated via MZMs is robust against inhomogeneous potentials, which inevitably exist at junctions, and also against disorder potentials. These results imply that detecting Majorana teleportation through non-local conductance measurements can serve as solid evidence for the existence of MZMs. 

This work was supported by JST CREST Grant No. JPMJCR19T5, Japan, a Grant-in-Aid for Scientific Research on Innovative Areas, Quantum Liquid Crystals (JP22H04480) from JSPS of Japan, and JSPS KAKENHI (Grants No. JP20K03860, No. JP20H01857, No. JP21H01039, and No. JP22H01221).

\bibliographystyle{jpsj}  
\bibliography{JPSJ}  

\end{document}


\maketitle
\begin{figure*}
  \includegraphics[width=\columnwidth]{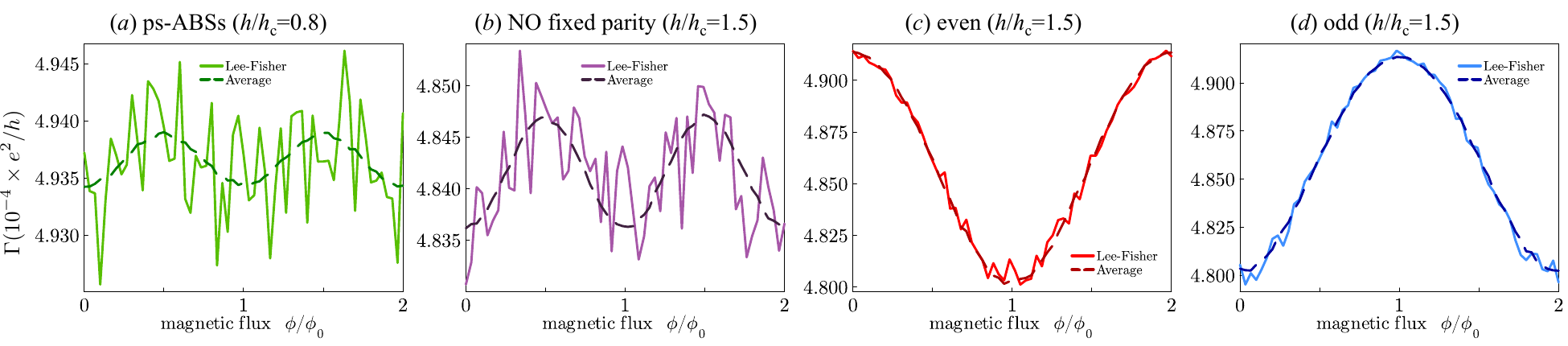}
  \caption{
    Magnetic flux dependence of the non-local conductance under disorder potentials for ps-ABSs with $h/h_{c}=0.8$ (a), MZMs with $h/h_{c}=1.5$  without fixed parity (b), and with parity even (\textit c) and odd (\textit d).
    The solid lines represent the result for a single configuration of disorder potentials, and the dashed lines represent the averages over impurity distributions. 
  }
  \label{fig6}
\end{figure*}
\begin{figure*}
  \includegraphics[width=\columnwidth]{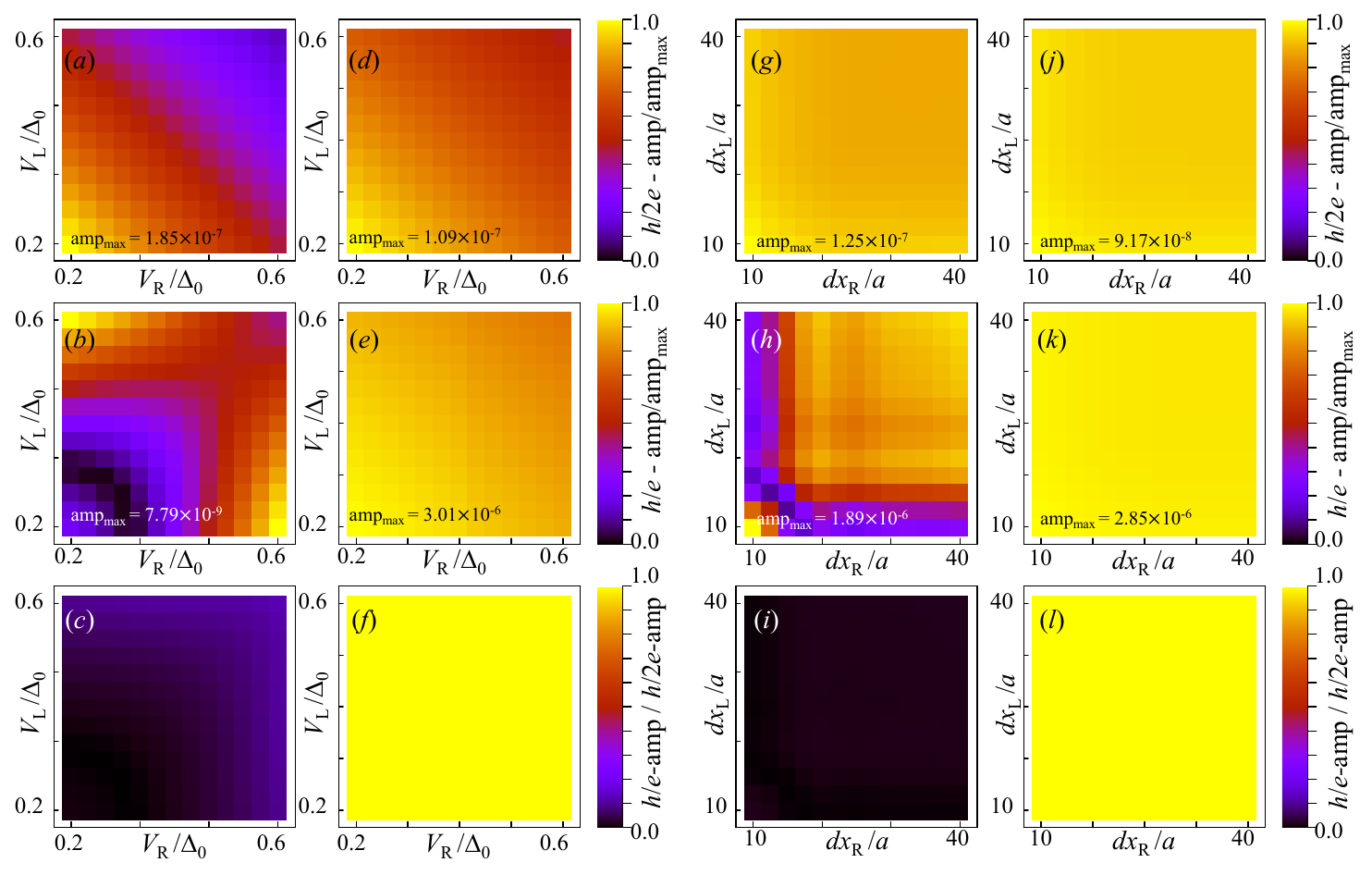}
  \caption{
    The amplitudes of the Fourier-transformed non-local conductance under disorder potential as functions of potential height (\textit a-\textit f) and width (\textit g-\textit l) at the left and right ends.
    The top, middle, and bottom rows show the amplitudes with $h/2e$-periodicity, the amplitudes with $h/e$-periodicity, and 
    the ratio of the $h/e$-amplitude to the $h/2e$-amplitude, respectively.
    In  (\textit a-\textit c) and (\textit g-\textit i), the results for the trivail phase with $h/h_{c}=0.8$ are shown. In (\textit d-\textit f) and (\textit j-\textit l), the results for the topological phase with $h/h_{c}=1.5$ are shown.
    }
  \label{fig7}
\end{figure*}

We examine effects of disorder potentials in the nanowire system on Majorana teleportation interference.
The Hamiltonian for disorder potentials is given by,
\begin{align}
  \mathcal{H}_{\rm{impu}} = \sum_{j,j',\sigma,\sigma'} \left[V_{j}c_{j,\sigma}^{\dagger} c_{j,\sigma} + V_{j'}\psi_{j',\sigma'}^{\dagger} \psi_{j',\sigma'} \right],
\end{align}
where $V_{j}$ is a random potential at site $j$, which is assigned a random number within the range $|V_j|\leq10^{-4}t$.
In one-dimensional electron systems, random potentials generally give rise to Anderson localization. However, for this range of random potentials, the localization length is larger than the length of the SN-loop system.
Thus, we can avoid complexity arising from Anderson localization. 
Figure~\ref{fig6} shows the magnetic flux dependence of the non-local conductance in the case with impurities. 
The non-local conductance shows $h/e$-periodicity in the topological phase with fixed parity  [Fig.\ref{fig6}(\textit c,\textit d)], which implies that Majorana teleportation can occur even in the presence of disorder potentials in addition to inhomogeneous potentials.
Moreover, we investigate the stability of Majorana teleportation by varying the height and width of the inhomogeneous potentials. 
The Majorana teleportation is stable also against the change of the potential structures even in the case with impurities, as seen in Fig.~\ref{fig7}.
The $h/e$-periodicity characterizing Majorana teleportation is stable against disorder potentials as well as inhomogeneous potentials at junctions.

